\documentclass[12pt]{article}

\begin{document}

\begin{center}
{\bf Black hole radiation of spin-1 particles in (1+2) dimensions} \\
\vspace{5mm} S. I. Kruglov
\footnote{E-mail: serguei.krouglov@utoronto.ca}

\vspace{3mm}
\textit{Department of Chemical and Physical Sciences, University of Toronto,\\
3359 Mississauga Road North, Mississauga, Ontario L5L 1C6, Canada} \\
\vspace{5mm}
\end{center}

\begin{abstract}
The radiation of vector particles by black holes in (1+2) dimensions is investigated within the WKB approximation. We
consider the process of quantum tunnelling of bosons through an event horizon of the black hole. The emission temperature
for the Schwarzschild background geometry coincides with the Hawking temperature and for the Rindler spacetime the
temperature is the Unruh temperature. We also obtain the radiation temperature for the de Sitter spacetime.
\end{abstract}

The black hole radiation of scalar particles was investigated by Hawking based on semiclassical calculations
using the Wick Rotation method \cite{Hawking}, \cite{Hawking'}, \cite{Hawking''}. With the help of quantum mechanics
the thermal radiation was studied
which is due to a horizon of the Schwarzschild spacetime background.
Without affecting the background, tunnelling scalar particles through the spherically symmetric Schwarzschild black
holes was investigated in \cite{Hawking1}. The Hawking radiation was also considered as a quantum tunnelling effect near
the horizon in \cite{Wilczek}, \cite{Wilczek'}, \cite{Wilczek1}, \cite{Shankaranarayanan}, \cite{Shankaranarayanan1}, \cite{Shankaranarayanan2} with the help of the
WKB approximation. The Unruh temperature \cite{Unruh} was obtained for a Rindler spacetime in \cite{Padmanabhan},
\cite{Mann}, \cite{Mann1}. It was shown that spin-1/2 and spin-3/2 particles and photons can also be produced by black
holes at the Hawking temperature \cite{Mann}, \cite{Mann1}, \cite{Mann3}, \cite{Jing}, \cite{Majhi}, \cite{Yale} using
the WKB approximation at leading order in the Planck constant $\hbar$. The emission and absorption probabilities are
connected by the relation
\begin{equation}
P[\mbox{emission}]=P[\mbox{absorption}]\exp\left(-\beta E\right),
\label{1}
\end{equation}
where $\beta$ is the inverse temperature and $E$ is an energy. Thus, to calculate the temperature we should
evaluate the emission and absorption probabilities for the processes through the horizon. It should be noted that
there is no barrier for a particle to fall into a black hole, and the barrier exists only for
the outgoing particles. Therefore, the absorption probability has to be normalized in such a way that
$P[\mbox{absorption}]=1$.
The emission spectrum of a black hole radiation is similar to a black body radiation \cite{Page}, \cite{Page'}.
A black hole may radiate arbitrary spin particles and, therefore, it is interesting to applied the tunnelling method for
the emission of spin-1 particles (spin-1 bosons $Z$ and $W^\pm$ are well-known).
We consider the radiation of spin-1 bosons within the Proca equations in (1+2) dimensions. The probability
of pair production of arbitrary spin particles by the electric field was considered in \cite{Kruglov}, \cite{Kruglov'},
\cite{Kruglov''}.

The system of units $c = G = k_B = 1$ is used.

Let us consider the polar metric in $(1+2)$ dimensions with the line element:
\begin{equation}
ds^2=-A(r)dt^2+\frac{1}{B(r)}dr^2+C(r)d\varphi^2.
\label{2}
\end{equation}
When the $B(r)$ vanishes ($B(r_0)=0$) the point $r = r_0$ defines the horizon in this spacetime. We
have $B(r)=\partial_rB(r_0)+{\cal O}[(r-r_0)^2]$ in the point which is close to the horizon.
Vector particles are described by the Proca equations that are given by
\[
D_\mu\psi^{\nu\mu}+\frac{m^2}{\hbar^2}\psi^\nu=0,
\]
\vspace{-8mm}
\begin{equation}
\label{3}
\end{equation}
\vspace{-8mm}
\[
\psi_{\nu\mu}=D_\nu\psi_\mu-D_\mu\psi_\nu=\partial_\nu\psi_\mu-\partial_\mu\psi_\nu,
\]
where $D_\mu$ are covariant derivatives, and $\psi_\nu=(\psi_0,\psi_1, \psi_2)$.
Different properties of the Proca equations in the flat spacetime are described in \cite{Kruglov2}.
With the help of the equation \cite{Landau}
\begin{equation}
D_\mu\psi^{\nu\mu}=\frac{1}{\sqrt{-g}}\frac{\partial\left(\sqrt{-g}\psi^{\nu\mu}\right)}{\partial x^\mu},
\label{4}
\end{equation}
one obtains from Eqs.(3) for the metric (2), $g_{\mu\nu}=\mbox{diag}(-A(r),1/B(r),C(r))$, $g=-A(r)C(r)/B(r)$, the system
of three equations
\[
\sqrt{\frac{AB}{C}}\left\{\partial_r\left[\sqrt{\frac{BC}{A}}\left(\partial_t\psi_1-\partial_r\psi_0\right)
\right]+\partial_\varphi\left[\frac{1}{\sqrt{ABC}}\left(\partial_t\psi_2-\partial_\varphi\psi_0\right)\right]\right\}
+ \frac{m^2}{\hbar^2}\psi_0=0,
\]
\begin{equation}
\frac{1}{\sqrt{ABC}}\left\{\partial_t\left[\sqrt{\frac{BC}{A}}\left(\partial_t\psi_1-\partial_r\psi_0\right)
\right]+\partial_\varphi\left[\sqrt{\frac{AB}{C}}\left(\partial_r\psi_2-\partial_\varphi\psi_1\right)\right]\right\}
+ \frac{m^2}{\hbar^2}\psi_1=0,
\label{5}
\end{equation}
\[
\sqrt{\frac{BC}{A}}\left\{\partial_t\left[\frac{1}{\sqrt{ABC}}\left(\partial_t\psi_2-\partial_\varphi\psi_0\right)
\right]+\partial_r\left[\sqrt{\frac{AB}{C}}\left(\partial_\varphi\psi_1-\partial_r\psi_2\right)\right]\right\}
+ \frac{m^2}{\hbar^2}\psi_2=0.
\]
We have used the covariant components $\psi_0=-A\psi^0$, $\psi_1=(1/B)\psi^1$, $\psi_2=C\psi^2$.
The solution to Eqs.(5) exists in the form
\begin{equation}
\psi_\nu=\left(c_0,c_1,c_2\right)\exp\left(\frac{i}{\hbar}S(t,r)\right).
\label{6}
\end{equation}
With the help of the WKB approximation, one can represent the action as follows:
\begin{equation}
S(t,r)=S_0(t,r)+\hbar S_1(t,r)+\hbar^2S_2(t,r)+....
\label{7}
\end{equation}
Eqs.(5), taking into account Eqs.(6),(7), at the leading order in $\hbar$ become
\[
B\left[c_0(\partial_rS_0)^2-c_1(\partial_rS_0)(\partial_tS_0)\right]+
\frac{1}{C}\left[c_0(\partial_\varphi S_0)^2-c_2(\partial_\varphi S_0)(\partial_tS_0)\right]+m^2c_0=0,
\]
\begin{equation}
\frac{1}{A}\left[c_0(\partial_rS_0)(\partial_tS_0)-c_1(\partial_tS_0)^2\right]+
\frac{1}{C}\left[c_1(\partial_\varphi S_0)^2-c_2(\partial_\varphi S_0)(\partial_r S_0)\right]+m^2c_1=0,
\label{8}
\end{equation}
\[
\frac{1}{A}\left[c_0(\partial_\varphi S_0)(\partial_tS_0)-c_2(\partial_tS_0)^2\right]+
B\left[c_2(\partial_r S_0)^2-c_1(\partial_\varphi S_0)(\partial_r S_0)\right]+m^2c_2=0.
\]
We look for the solution to Eqs.(8) in the form
\begin{equation}
S_0=-Et+W(r)+J(\varphi)+K,
\label{9}
\end{equation}
where $E$ is an energy, and $K$ is a complex constant. Putting Eq.(9) into Eqs.(8) one obtains the matrix equation
$\Lambda\left( c_1 , c_2 , c_3\right)^T=0$ (the superscript $T$ means the transition to the transposed vector), where
\begin{equation}
\Lambda=\left(
\begin{array}{ccc}
  B(\partial_rW)^2+\frac{(\partial_\varphi J)^2}{C}+m^2 & BE\partial_rW & \frac{E\partial_\varphi J}{C} \\
  \frac{E\partial_rW}{A} & \frac{E^2}{A}-\frac{(\partial_\varphi J)^2}{C}-m^2 & \frac{(\partial_rW)\partial_\varphi J}{C} \\
  \frac{E\partial_\varphi J}{A} & B(\partial_rW)\partial_\varphi J & \frac{E^2}{A}-B(\partial_rW)^2-m^2
\end{array}
\right).
\label{10}
\end{equation}
The system of linear equations $\Lambda\left( c_1 , c_2 , c_3\right)^T=0$ possesses nontrivial solutions if
$\mbox{det}\Lambda=0$. Then from equations (10) we obtain
\begin{equation}
\mbox{det}\Lambda=\left[B(\partial_rW)^2-\left(\frac{E^2}{A}-\frac{(\partial_\varphi J)^2}{C}-m^2\right)\right]^2,
\label{11}
\end{equation}
and the requirement $\mbox{det}\Lambda=0$ leads to the function
\begin{equation}
W_\pm(r)=\pm \int\sqrt{\frac{E^2-A(r)\left(m^2+\frac{(\partial_\varphi J)^2}{C(r)}\right)}{A(r)B(r)}}dr.
\label{12}
\end{equation}
For the outgoing motion of particles we have to use $W_+$ in Eq.(12) (the momentum is $p_r=\partial_rS_0>0$) and for
the ingoing motion - $W_-$ (the momentum is $p_r=\partial_rS_0<0$). The evaporation of black holes is a quantum
process \cite{Hawking} and
the emission and absorbtion probabilities of crossing the horizon out and in are
given by \cite{Mann1}
\begin{equation}
P[\mbox{emission}]=\exp\left[-\frac{2}{\hbar}\left(\mbox{Im}W_++\mbox{Im}K\right)\right],
\label{13}
\end{equation}
\begin{equation}
P[\mbox{absorption}]=\exp\left[-\frac{2}{\hbar}\left(\mbox{Im}W_-+\mbox{Im}K\right)\right].
\label{14}
\end{equation}
Since the $W_\pm(r)$ possesses a simple pole at the horizon $r=r_0$, we can use the complex contour and for outgoing
particles to make a replacement $r_0\rightarrow r_0-i\varepsilon$ \cite{Padmanabhan}.
For an absorbtion the probability should be normalized by choosing the imaginary part of the constant $K$ to be
Im$K$=-Im$W_-$ \cite{Mann1}. In this case there is not a reflection and the absorption probability in Eq.(14) is unity,
$P[\mbox{absorption}]=1$. Because $W_-=-W_+$, we have Im$K$=Im$W_+$, and as a result, from Eq.(13) we obtain
\begin{equation}
P[\mbox{emission}]=\exp\left(-\frac{4}{\hbar}\mbox{Im}W_+ \right).
\label{15}
\end{equation}
To evaluate the imaginary part of the integral (12) we use the formula \cite{Bogolubov}
\begin{equation}
\frac{1}{r-i\varepsilon}=i\pi\delta(r)+{\cal P}\left(\frac{1}{r}\right),
\label{16}
\end{equation}
where ${\cal P}\left(\frac{1}{r}\right)$ is the principal value of $1/r$.
Then the probability of tunnelling particles from inside to outside of the black hole becomes
\begin{equation}
P=\exp\left(-\frac{4\pi E}{\hbar\sqrt{\left(\partial_rA(r_0)\right)
\partial_rB(r_0)}}+ {\cal O}(E^2)\right).
\label{17}
\end{equation}
Here we imply that the function $C(r_0)$ is finite. Since Eq.(17) is similar to the Boltzmann formula (1)
for the particle emission, we obtain the temperature which is associated with the horizon
\begin{equation}
T=\frac{\hbar\sqrt{\left(\partial_rA(r_0)\right)\partial_rB(r_0)}}{4\pi }.
\label{18}
\end{equation}
The expression like (18) was obtained also for the black hole emission of particles with spins: $0$, $1/2$, $3/2$
\cite{Mann}, \cite{Mann1}, \cite{Mann3}, \cite{Jing}.
We can use $A(r)=B(r)=1-2M/r$ for the Schwarzschild black hole metric, where $M$ is a mass of the black hole
(the event horizon is at $r_0=2M$). The radiation temperature, obtained from Eq.(18) for an observer at $l >2M$,
becomes $T_H=\hbar/(8\pi M)$ that is the Hawking temperature.

If $A(r)=B(r)=1-H^2r^2$ with $H=\dot{a}(t)/a(t)$ being the Hubble parameter, one has the de Sitter spacetime.
The event horizon is given by $r_0=1/H$, and from Eq.(18) we obtain the temperature $T=\hbar H/(2\pi)$.

One can use the functions $A(r)=B(r)=1+2gr$ ($g$ is an acceleration) for the Rindler spacetime.
Then from Eq.(18) we recover the Unruh temperature $T=\hbar g/(2\pi)$.

Thus, the quantum tunnelling method, due to the presence of the horizon, allows us to recover the Hawking temperature
for black holes emission of vector bosons. It should be noted that the tunnelling method is simpler compared with
the Bogolyubov transformation method \cite{Birrel} and can be used for different metrics and spin particles.
One can calculate non-thermal corrections to the emission temperature by taking into account higher orders in $\hbar$.
As a result, the spectrum will not be precisely thermal because of the $E^2$ corrections to the Boltzmann expression.
The generalization of results obtained on the ($1+3$) dimension is straightforward.

\end{document}